\documentclass[11pt]{article}
\usepackage{amssymb}
\begin{document}
\setlength{\oddsidemargin}{0cm} \setlength{\evensidemargin}{0cm}
\baselineskip=20pt

\begin{center} {\Large\bf
Some Non-Abelian Phase Spaces in Low Dimensions}
\end{center}

\bigskip

\begin{center}  { \large Dongping Hou\qquad \large Chengming Bai}\footnote{Corresponding author. E-mail address:
baicm@nankai.edu.cn}  \end{center}

\begin{center}{\it Chern Institute of Mathematics \& LPMC, Nankai University, Tianjin
300071, P.R. China}\end{center}

\bigskip

\centerline{\large\bf   Abstract }
\vspace{0.2cm}

A non-abelian phase space, or a phase space of a Lie algebra is a
generalization of the usual (abelian) phase space of a vector space.
It corresponds to a parak\"ahler structure in geometry. Its
structure can be interpreted in terms of left-symmetric algebras. In
particular, a solution of an algebraic equation in a left-symmetric
algebra which is an analogue of classical Yang-Baxter equation in a
Lie algebra can induce a phase space. In this paper, we find that
such phase spaces have a symplectically isomorphic property. We also
give all such phase spaces in dimension 4 and some examples in
dimension 6. These examples can be a guide for a further
development.

\vspace{0.5cm}


\vspace{0.5cm}


\newpage

\section{Introduction}

It is known that the phase space $T^*V$ of a vector space $V$ over a
field ${\bf F}$ can be defined as the direct sum of $V$ and its dual
space $V^*={\rm Hom}(V, {\bf F})$ endowed with the symplectic form
$$\omega_p (x+a^*,y+b^*)=\langle a^*,y\rangle -\langle b^*,x\rangle ,\;\;\forall x,y\in V,a^*,b^*\in V^*, \eqno (1.1)$$
where $\langle ,\rangle $ is the ordinary pairing between $V$ and
$V^*$. In [Ku1], Kupershmidt generalized the above definition to the
non-abelian case in the sense of replacing $V$ by  a Lie algebra (in
particular for a non-abelian Lie algebra). Let ${\cal G}$ be a Lie
algebra and ${\cal G}^*$ be its dual space. A phase space of ${\cal
G}$ is the vector space $T^*({\cal G})={\cal G}\oplus {\cal G}^*$ as
the direct sum of vector spaces such that $T^*({\cal G})$ is a Lie
algebra, ${\cal G}$ is its subalgebra and the symplectic form
$\omega_p$ given by equation (1.1) is a 2-cocycle on $T^*({\cal
G})$, that is,
$$\omega_p([x_1+a_1^*, x_2+a_2^*], x_3+a_3^*)+{\rm CP}=0,\;\;\forall x_i\in {\cal
G},\;a_i^*\in {\cal G}^*,\eqno (1.2)$$ where ``CP" stands for
``cyclic permutation". A further study of non-abelian phase spaces
was given in [Ba2]. In particular, the module structure on ${\cal
G}^*$ in [Ku1] (it is equivalent to the condition that ${\cal G}^*$
is an ideal of $T^*({\cal G})$) can be generalized to be a
subalgebra. However, it is not easy to get some concrete examples
under this sense.

On the other hand, the (non-abelian) phases spaces are just the
parak\"ahler structures on Lie algebras. In geometry, a parak\"ahler
manifold is a symplectic manifold with a pair of transversal
Lagrangian foliations ([L]). The Lie algebra ${\cal G}$ of a Lie
group $G$ with a $G$-invariant parak\"ahler structure is a
parak\"ahler structure on ${\cal G}$ ([BMO], [BBM], [MR3]). It is a
symplectic Lie algebra ([C], [LM], [MS], etc.) which is a direct sum
of the underlying vector spaces of two Lagrangian subalgebras
([Ka]).

In fact, both of these two structures can be interpreted in terms of
a kind of nonassociative algebras, namely, left-symmetric algebras
(or Koszul-Vinberg algebras). Left-symmetric algebras arose from the
study of convex homogeneous cones, affine manifolds and deformation
of algebras ([V], [G], [A], [Ki], [M]) and appeared in many fields
in mathematics and mathematical physics, such as complex and
symplectic structures on Lie groups and Lie algebras ([H], [C],
[DaM1-2], [KGN], [AS]), integrable systems ([SS]), classical and
quantum Yang-Baxter equations ([Bo], [ES], [Ku2-3], [GS], [DiM]),
Poisson brackets and infinite-dimensional Lie algebras ([GD],
[BN],[Z]), vertex algebras ([BK]), quantum field theory ([CK]),
operads ([CL]) and so on.

In [Ba3], we have known that a phase space is isomorphic to a
bialgebra structure, namely, a left-symmetric bialgebra. It has many
similar properties of a Lie bialgebra ([D]). In particular, such a
structure (hence the phase space) can be obtained through solving an
algebraic equation ($S$-equation) in left-symmetric algebras which
is an analogue of classical Yang-Baxter equation in Lie algebras
([Se], [BD], [Ku3]).

In this paper, we give a further study of $S$-equation. We find a
symplectically isomorphic property of the phase spaces constructed
through $S$-equation. We also give such phases in low dimensions and
we hope that these examples can be a guide for a further
development. This paper is organized as follows. In section 2, for
self-contained, we give a brief introduction to phase spaces and
left-symmetric algebras. In section 3, we recall the construction of
phase spaces through $S$-equation in left-symmetric algebras given
in [Ba3] and we prove a symplectically isomorphic property of such
phase spaces. In section 4, we give the 4-dimensional phase spaces
obtained from solving $S$-equation in 2-dimensional complex
left-symmetric algebras. In section 5, we give some examples in
dimension 6 through giving all solutions of $S$-equation in
3-dimensional complex simple left-symmetric algebras. In section 6,
we give some conclusion and discussion.

Throughout this paper, all algebras are finite dimensional and over
the complex field $\bf C$ and the parameters belong to the complex
field $\bf C$, too. And $\langle  | \rangle $ stands for a Lie or
left-symmetric algebra with a basis and nonzero products at each
side of $|$.

\section{Preliminaries and some basic results}

\mbox{}\hspace{0.6cm} {\bf Definition 2.1}\quad A parak\"ahler Lie
algebra ${\cal G}$ or a parak\"ahler structure on a Lie algebra
${\cal G}$ is a triple $\{{\cal G}^{+},{\cal G}^{-},\omega\}$, where
${\cal G}^{+},{\cal G}^{-}$ are two subalgebras of $\cal G$ and
${\cal G}={\cal G}^{+}\oplus{\cal G}^{-}$ as vector spaces, $\omega$
(the symplectic form) is a nondegenerate skew-symmetric 2-cocycle on
$\cal G$ such that
$$\omega([x,y],z)+\omega([y,z],x)+\omega ([z,x],y)=0,\;\;\forall
x,y,z\in {\cal G},\eqno (2.1)$$ and $\omega({\cal G}^{+},{\cal
G}^{+})=\omega({\cal G}^{-},{\cal G}^{-})=0$.

{\bf Definition 2.2}\quad Two parak\"ahler Lie algebras $({\cal
 G}_1^+,{{\cal G}_1 }^{-},\omega_1)$ and $({\cal
 G}_2^+,{{\cal G}_2 }^{-},\omega_2)$ are isomorphic if there exists
 a Lie algebra isomorphism $\varphi:{\cal G}_1 \rightarrow
{\cal G}_2$ such that
$$\varphi({\cal G}_1^+)={\cal G}_2^+,\; \varphi({\cal G}_1^-)={\cal G}_2^-;
\;\;\omega_1(x,y)=\varphi^*\omega_2(x,y)=\omega_2(\varphi(x),\varphi(y)),\;\forall
x,y\in {\cal G}_1. \eqno (2.2)$$

{\bf Proposition 2.3 }([Ba3])\quad  Any Lie algebra ${\cal G}$'s
phase space $T^*({\cal G})$ admits a parak\"ahler structure $({\cal
G}, {\cal G}^*, \omega_p)$, where $\omega_p$ is given by equation
(1.1). Conversely, every parak\"ahler Lie algebra $({\cal G}^+,{\cal
G}^-,\omega)$ is isomorphic to a phase space of ${\cal G}^+$.

%On the other hand,

{\bf Definition 2.4}\quad Let $A$ be a vector space over a filed
{\bf F} with a bilinear product $(x,y)\rightarrow xy$. $A$ is called
a left-symmetric algebra, if for any $x,y,z\in A$,
$$(xy)z-x(yz)=(yx)z-y(xz).\eqno (2.3)$$

{\bf Proposition 2.5} ([M])\quad Let $A$ be left-symmetric algebra.
For any $x\in A$, let $L_x$ denote the left multiplication operator,
that is, $L_x(y)=xy,\;\forall\; y\in A$. Then we have

(i)\quad The commutator
$$[x,y]=xy-yx,\;\;\forall\; x,y\in A,\eqno (2.4)$$
defines a Lie algebra ${\cal G}={\cal G}(A)$, which is called the
sub-adjacent Lie algebra of $A$ and $A$ is  called the compatible
left-symmetric algebra structure on the Lie algebra ${\cal G}(A)$.

(ii)\quad
%From equation (2.3) and (2.4), we have
%$$[L_x, L_y]=L_{[x,y]},\;\;\forall\; x,y\in A,\eqno (2.5)$$
%that is,
$L:{\cal G}(A)\rightarrow gl({\cal G}(A))$ with
$x\rightarrow L_x$ gives a (regular) representation of the
sub-adjacent Lie algebra ${\cal G}(A)$.

%The phase spaces (parak\"ahler Lie algebras) can be interpreted in
%terms of left-symmetric algebras.

{\bf Theorem 2.6 ([C], [Ba2])}\quad Let $\{{\cal G}^+,{\cal
G}^-,\omega\}$ be a parak\"ahler structure on a Lie algebra $\cal
G$. Then there exists a compatible left-symmetric algebra structure
$``*"$ on $\cal G$ defined by
$$\omega(x*y,z)=-\omega(y,[x,z]),\;\;\forall x,y,z\in {\cal
G}.\eqno (2.5)$$ Moreover, ${\cal G}^{\pm}$ are two (left-symmetric)
subalgebras of ${\cal G}$ with the above product.

Let ${\cal G}$ be a Lie algebra and $\rho:{\cal G}\rightarrow gl(V)$
be a representation. On a direct sum ${\cal G}\oplus$ of the
underlying vector spaces ${\cal G}$ and $V$, there is a natural Lie
algebra structure (denoted by ${\cal G}\ltimes_\rho V$) given as
follows ([J]).
$$[x_1+v_1, x_2+v_2]=[x_1,x_2]+\rho (x_1)v_2-\rho (x_2)v_1,\;\forall
x_1,x_2\in {\cal G}, v_1, v_2\in V. \eqno (2.6)$$

{\bf Example 2.7}\quad Let $A$ be a left-symmetric algebra. Then
$T^*({\cal G})(A)={\cal G}(A)\ltimes_{L^*}{\cal G}(A)^*$ is a phase
space of the sub-adjacent Lie algebra ${\cal G}(A)$ with the
symplectic form $\omega_p$ given by equation (1.1), where ${\cal
G}(A)^*$ is the dual space of ${\cal G}(A)$ and $L^*$ is the dual
representation of the regular representation $L$ of ${\cal G}(A)$.
Such a construction was given by Medina and Revoy ([MR1-2]) and
Kupershmidt ([Ku1]) respectively.

{\bf Definition 2.8}\quad  Let $A$ be a vector space. A
left-symmetric bialgebra structure on $A$ is a pair of linear maps
$(\alpha,\beta)$ such that $\alpha:A\rightarrow A\otimes A$, $\beta:
A^*\rightarrow A^*\otimes A^*$ and

(a) $\alpha^*:A^*\otimes A^*\rightarrow A^*$ is a left-symmetric
algebra structure on $A^*$;

(b) $\beta^*:A\otimes A\rightarrow A$ is a left-symmetric algebra
structure on $A$;

(c) $\alpha$ is a 1-cocycle of ${\cal G}(A)$ associated to $L\otimes
1+1\otimes {\rm ad}$ with values in $A\otimes A$;

(d) $\beta$ is a 1-cocycle of ${\cal G}(A^*)$ associated to
$L\otimes 1+1\otimes {\rm ad}$ with values in $A^*\otimes A^*$;

\noindent where $L$ is the regular representation of ${\cal G}(A)$
and ${\rm ad}$ is the adjoint representation of ${\cal G}(A)$
satisfying ${\rm ad}x(y)=[x,y]=xy-yx$ for any $x,y\in A$. We also
denote this left-symmetric bialgebra by $(A,A^*,\alpha,\beta)$ or
simply $(A,A^*)$. Two left-symmetric bialgebras
$(A,A^*,\alpha_A,\beta_A)$ and $(B,B^*,\alpha_B,\beta_B)$ are
isomorphic if there is a left-symmetric algebra isomorphism
$\varphi:A\rightarrow B$ such that $\varphi^*:B^*\rightarrow A^*$ is
also an isomorphism of left-symmetric algebras, that is, $\varphi$
satisfies
$$(\varphi\otimes \varphi) \alpha_A (x)=\alpha_B(\varphi(x)),\;\;
(\varphi^*\otimes
\varphi^*)\beta_B(a^*)=\beta_A(\varphi^*(a^*)),\forall x\in A,a^*\in
B^*.\eqno (2.7)$$

{\bf Theorem 2.9} ([Ba3])\quad Let $(A,\cdot)$ be a left-symmetric
algebra and $(A^*,\circ)$ be a left-symmetric algebra structure on
its dual space $A^*$. Then $({\cal G}(A),{\cal G}(A)^*, \omega_p)$
(where $\omega_p$ is given by equation (1.1)) is a phase space if
and only if $(A,A^*)$ is a left-symmetric bialgebra. Moreover, two
phase spaces are isomorphic if and only if their corresponding
left-symmetric bialgebras are isomorphic.

\section{$S$-equation and symplectic isomorphism}

%There is a special case of left-symmetric bialgebras as follows.

\mbox{}\hspace{0.6cm}{\bf Definition 3.1}\quad A left-symmetric
bialgebra $(A,A^*,\alpha,\beta)$ is called coboundary if $\alpha$ is
a 1-coboundary of ${\cal G}(A)$ associated to $L\otimes 1+1\otimes
{\rm ad}$, that is, there exists a $r\in A\otimes A$ such that
$$\alpha (x)=(L_x\otimes 1+1\otimes {\rm ad}x)r,\;\;\forall x\in A.\eqno (3.1)$$

{\bf Notation 3.2}\quad Let $(A,\cdot)$ be a left-symmetric algebra
and $r=\sum_i a_i\otimes b_i\in A\otimes A$. Set
$$r_{12}=\sum_ia_i\otimes b_i\otimes 1,\;r_{13}=\sum_{i}a_i\otimes 1\otimes b_i;\;r_{23}=\sum_i1\otimes
a_i\otimes b_i\;\in U({\cal G}(A)),\eqno (3.2)$$ where $U({\cal
G}(A))$ is the universal enveloping algebra of the sub-adjacent Lie
algebra ${\cal G}(A)$. Set
$$r_{12}\cdot r_{13}=\sum_{i,j}a_i\cdot
a_j\otimes b_i\otimes b_j;\;\; r_{12}\cdot r_{23}=\sum_{i,j}
a_i\otimes b_i\cdot a_j\otimes b_j;\eqno (3.3)$$
$$[r_{13},r_{23}]=r_{13}\cdot r_{23}-r_{23}\cdot
r_{13}=\sum_{ij}a_i\otimes a_j\otimes [b_i,b_j].\eqno (3.4)$$

{\bf Proposition 3.3}\quad ([Ba3])\quad Let $A$ be a left-symmetric
algebra and $r\in A\otimes A$. Suppose $r$ is symmetric. Then the
map $\alpha$ defined by equation (3.1) induces a left-symmetric
algebra structure on $A^*$ such that $(A,A^*)$ is a left-symmetric
bialgebra if
$$[[r,r]]=-r_{12}\cdot r_{13}+r_{12}\cdot r_{23}+[r_{13},r_{23}]=0.\eqno (3.5)$$

Let $A$ be a vector space. For any $r\in A\otimes A$, $r$ can be
regarded as a map from $A^*\rightarrow A$ in the following way:
$$\langle u^*\otimes v^*,r\rangle =\langle u^*,r(v^*)\rangle ,\;\;\forall u^*,v^*\in A^*.\eqno (3.6)$$
 $r$ is symmetric if and only if $\langle u^*\otimes v^*,
r\rangle =\langle v^*\otimes u^*, r\rangle $ for any $u^*,v^*\in
A^*$.

{\bf Remark 3.4}\quad Equation (3.5) is called a $S$-equation in a
left-symmetric algebra. This algebraic equation has a geometric
meaning as follows. Let $(A,\cdot)$ be a left-symmetric algebra and
a bilinear form ${\cal B}:A\otimes A\rightarrow {\bf F}$ is called a
2-cocycle of $A$ if
$${\cal B}(x\cdot y,z)-{\cal B}(x,y\cdot z)={\cal B}(y\cdot x,z)-{\cal B}(y,x\cdot z),\forall x,y,z\in A.\eqno (3.7)$$
Suppose that $r\in A\otimes A$ is symmetric and nondegenerate. Then
$r$ is a solution of $S$-equation in $A$ if and only if the inverse
of the isomorphism $A^*\rightarrow A$ induced by $r$, regarded as a
bilinear form ${\cal B}$ on $A$, is a 2-cocycle of $A$. That is,
${\cal B}(x,y)=\langle r^{-1}x,y\rangle $ for any $x,y\in A$.

On the other hand, a left-symmetric algebra over the real field
${\bf R}$ is called Hessian if there exists a symmetric and positive
definite 2-cocycle ${\cal B}$ of $A$. In geometry, a Hessian
manifold $M$ is a flat affine manifold provided with a Hessian
metric $g$, that is, $g$ is a Riemannian metric such that for any
each point $p\in M$ there exists a $C^\infty$-function $\varphi$
defined on a neighborhood of $p$ such that
$g_{ij}=\frac{\partial^2\varphi}{\partial x^i\partial x^j}$. A
Hessian left-symmetric algebra corresponds to an affine Lie group
$G$ with a $G$-invariant Hessian metric ([Sh]). \hfill $\Box$

{\bf Proposition 3.5}\quad ([Ba3])\quad Let $(A,\cdot)$ be a
left-symmetric algebra and $r\in A\otimes A$ be a symmetric solution
of $S$-equation in $A$. Then the left-symmetric algebra and its
sub-adjacent Lie algebra structure in the phase space $T^*({\cal
G}(A))$ can be given from the products in $A$ as follows.

(a) $a^**b^*=a^*\circ b^*=-R^*_\cdot (r(b^*))a^*+{\rm
ad}_\cdot^*(r(a^*))b^*$, for any $a^*,b^*\in A^*$;\hfill (3.8)

(b) $[a^*,b^*]=a^*\circ b^*-b^*\circ a^*=L^*_\cdot
(r(a^*))b^*-L_\cdot^*(r(b^*))a^*$, for any $a^*,b^*\in A^*$;\hfill
(3.9)

(c) $x*a^*=x\cdot r(a^*)-r({\rm ad}_\cdot^*(x)a^*) + {\rm
ad}_\cdot^* (x)a^*$, for any $x\in A$, $a^*\in A^*$;\hfill (3.10)

(d) $a^**x=r(a^*)\cdot x+r(R_\cdot^* (x) a^*)-R_\cdot^*(x)a^*$, for
any $x\in A$, $a^*\in A^*$;\hfill (3.11)

(e) $[x,a^*]=[x,r(a^*)]-r(L_\cdot^*(x)a^*)+L_\cdot^*(x)a^*$, for any
$x\in A$, $a^*\in A^*$.\hfill (3.12)

\noindent where $R^*_\cdot: A\rightarrow gl(A^*)$ satisfying
$\langle R^*_\cdot (x)a^*,y\rangle =-\langle a^*,y\cdot x\rangle $
for any $x,y\in A$ and $a^*\in A^*$.

% Therefore, we can construct the
%(non-abelian) phase spaces through solving the $S$-equation in
%left-symmetric algebras and the structures of such phase spaces are
%given as above.

{\bf Example 3.6}\quad Obviously, $r=0$ is a symmetric solution of
$S$-equation in any left-symmetric algebra $A$. Moreover, it is
obvious that the phase space constructed from $r=0$ just corresponds
to the phase space in Example 2.7 as the semi-direct sum ${\cal
G}(A)\ltimes_{L^*}{\cal G}(A)^*$. \mbox{}\hfill $\Box$

%Furthermore, we have the following symplectically isomorphic
%property.

{\bf Definition 3.7}\quad A symplectic Lie algebra or a symplectic
structure on a Lie algebra is a pair $({\cal G}, \omega)$, where
${\cal G}$ is a Lie algebra and $\omega$ is a nondegenerate
skew-symmetric 2-cocycle on $\cal G$ satisfying equation (2.1). Two
symplectic Lie algebras $({\cal G}_1,\omega_1)$ and $({\cal
G}_2,\omega_2)$ are isomorphic if there exists a Lie algebra
isomorphism $\varphi:{\cal G}_1 \rightarrow {\cal G}_2$ such that
$$\omega_1(x,y)=\varphi^*\omega_2(x,y)=\omega_2(\varphi(x),\varphi(y)),\;\;\forall
x,y\in {\cal G}_1. \eqno (3.13)$$

%It is obvious that a phase space is a symplectic Lie algebra and an
%isomorphism of phase spaces is naturally an symplectic isomorphism.

{\bf Lemma 3.8} ([Ba3])\quad  Let $(A,\cdot)$ be a left-symmetric
algebra and $r\in A\otimes A$ be symmetric. Then $r$ is a solution
of $S$-equation in $A$ if and only if $r$ satisfies
$$[r(a^*),r(b^*)]=r(L_\cdot^*(r(a^*))b^*-L_\cdot^*(r(b^*))a^*),\;\;\forall a^*,b^*\in A^*.\eqno (3.14)$$

{\bf Theorem 3.9}\quad Let $A$ be a left-symmetric algebra. Then as
symplectic Lie algebras, the phase space $T^*({\cal G}(A))$ given by
any symmetric solution of $S$-equation in $A$ is isomorphic to the
phase space given in Example 2.7, that is, the semi-direct sum
${\cal G}(A)\ltimes_{L^*}{\cal G}(A)^*$ which corresponds to the
solution $r=0$.

{\bf Proof}\quad Let $r$ be a symmetric solution of $S$-equation in
$A$. Define a linear map $\varphi:{\cal G}(A)\ltimes_{L^*} {\cal
G}(A)^*\rightarrow A\oplus A^*$ satisfying
$$\varphi(x)=x,\;\;\varphi(a^*)=-r(a^*)+a^*,\;\;\forall x\in A, a^*\in A^*.$$
Obviously, $\varphi$ is a linear isomorphism. Since $r$ is symmetric
and by Lemma 3.8, we know

{\small
\begin{eqnarray*}
&&[\varphi(x),\varphi(y)]=[x,y]=\varphi([x,y]);\\
&&[\varphi(x),\varphi(a^*)]=[x,-r(a^*)+a^*]=-[x,r(a^*)]+[x,r(a^*)]-r(L^*(x)a^*)+L^*(x)a^*\\
&&\hspace{2.2cm}=-r(L^*(x)a^*)+L^*(x)a^*=\varphi(L^*(x)a^*)=\varphi([x,a^*]);\\
&&[\varphi(a^*),\varphi(b^*)]=
%[-r(a^*)+a^*, -r(b^*)+b^*]\\
%%&&\hspace{2.3cm}=
[r(a^*),r(b^*)]-\{
[r(a^*),r(b^*)]-r(L^*(r(a^*))b^*)+L^*(r(a^*))b^*\}
 \\
&&\hspace{2.3cm} +
\{[r(b^*),r(a^*)]-r(L^*(r(b^*))a^*)+L^*(r(b^*))a^*\}
+L^*(r(a^*))b^*-L^*(r(b^*))a^*\\
&&\hspace{2.3cm}=[r(b^*),r(a^*)]+r(L^*(r(a^*))b^*)-r(L^*(r(b^*))a^*)=0
=\varphi([a^*,b^*]);\\
&&\varphi^*\omega_p(x+a^*,y+b^*)=\langle a^*,y-r(b^*)\rangle -\langle x-r(a^*),b^*\rangle \\
&&\hspace{3.5cm}=\langle a^*,y\rangle -\langle x,b^*\rangle -\langle
a^*,r(b^*)\rangle +\langle r(a^*),b^*\rangle
%&&\hspace{3.5cm}=\langle a^*,y\rangle -\langle x,b^*\rangle
=\omega_p(x+a^*,y+b^*),
\end{eqnarray*}}
for any $x,y\in A$ and $a^*,b^*\in A^*$. Therefore $\varphi$ is an
isomorphism of symplectic Lie algebras. \hfill $\Box$

{\bf Remark 3.10}\quad In general, the above two phase spaces are
not isomorphic as parak\"ahler Lie algebras. In particular, for the
nonzero solution $r\ne 0$ of $S$-equation in $A$, the linear
isomorphism $\varphi$ given in the above proof is not an isomorphism
of parak\"ahler Lie algebras since $\varphi (A^*)=r(A^*)+A^*\ne A^*$
when $r\ne 0$. \hfill $\Box$

\section{4-dimensional phase spaces}

Let $\{e_1,\cdots,e_n\}$ be a basis of a left-symmetric algebra
$(A,\cdot)$ and $\{e_1^*,\cdots,e_n^*\}$ be its dual basis in $A^*$.
Let $r\in A\otimes A$. Set $r=\sum\limits_{i,j}^nr_{ij}e_i\otimes
e_j,\;\;e_i\cdot e_{j}=\sum\limits_{k=1}^n c_{ij}^k e_k$. Then
$r(e_i^*)=\sum\limits_{j=1}^n r_{ij}e_j$ and $r$ is a symmetric
solution of $S$-equation in $A$ if and only if $r_{ij}$ satisfies
%the following equations:
$$r_{ij}=r_{ji},\;\;\sum_{t,l}^n\{
-c_{tl}^ir_{tj}r_{lk}+c_{tl}^jr_{it}r_{lk}+(c_{tl}^k-c_{lt}^k)r_{it}r_{lj}\}=0,\;\;\forall
i,j,k=1,2,\cdots, n.\eqno (4.1)$$ We let ${\rm SE}(A)$ denote the
set of the symmetric solutions of $S$-equation in $A$. By
Proposition 3.5, we obtain the structures of phase spaces obtained
from solving the $S$-equation as follows.
\begin{eqnarray*}
&&[e_i,e_j] =\sum_{k=1}^n (c_{ij}^k-c_{ji}^k)e_k;\\
&&[e_i,e_j^*]=\sum_{k=1}^n[r_{jl}(c_{il}^k-c_{li}^k)
+\sum_{l=1}^nr_{lk}c_{il}^j]e_k-\sum_{k=1}^nc_{ik}^je_k^*;\\
&&[e_i^*,e_j^*]=\sum_{l,k}^n(-r_{il}c_{lk}^j+r_{jl}c_{lk}^i)e_k^*.
\end{eqnarray*}

Now we consider the case $n=2$. The classification of 2-dimensional
complex left-symmetric algebras (with non-zero products) are given
as follows ([Bu]).

\hspace {12pt}(AI)=$\langle e_1,e_2|e_1\cdot e_1=e_1,e_2\cdot
e_2=e_2\rangle ;$

\hspace {12pt}(AII)=$\langle e_1,e_2|e_2\cdot e_2=e_2,e_1\cdot
e_2=e_2\cdot e_1=e_1\rangle ;$

\hspace {12pt}(AIII)=$\langle e_1,e_2|e_1\cdot e_1=e_1\rangle $;

\hspace {12pt}(AV)=$\langle e_1,e_2|e_1\cdot e_1=e_2\rangle :\;$

\hspace {12pt}(AIV)=$\langle e_1,e_2|e_i\cdot e_j=0,i,j=1,2\rangle
;$

\hspace {12pt}(NI)=$\langle e_1,e_2|e_2\cdot e_1=-e_1,e_2\cdot
e_2=-e_2\rangle ;$

\hspace {12pt}(NII)$_{-1}=\langle e_1,e_2|e_2\cdot e_1=-e_1,e_2\cdot
e_2=e_1-e_2\rangle ;$

 \hspace {12pt}(NII)$_k=\langle e_1,e_2|e_2\cdot e_1=-e_1,e_2\cdot e_2=ke_2,k\ne {-1}\rangle ;$

 \hspace {12pt}(NIII)=$\langle e_1,e_2|e_1\cdot e_2=e_1,e_2\cdot e_2=e_2\rangle ;$

 \hspace
 {12pt}(NIV)$_k=\langle e_1,e_2|e_1\cdot e_2=ke_1,e_2\cdot e_1=(k-1)e_2,e_2\cdot e_2=e_1+ke_2,k\in
 {\bf C}\rangle ;$

 \hspace {12pt}(NV)=$\langle e_1,e_2|e_1\cdot e_1=2e_1,e_1\cdot e_2=e_2,e_2\cdot e_2=e_1\rangle .$

By a direct computation, we give the solutions of $S$-equation in
the above (2-dimensional) left-symmetric algebras and their
corresponding (4-dimensional) phase spaces (only the non-zero
products are given) as follows.

{\small \begin{eqnarray*}&&{\rm SE(AI)}=\{\left(\matrix{r_{11}&0\cr
0& r_{22}\cr}\right)\} \Longrightarrow \left\{\matrix{[e_1,
e_1^*]=r_{11}e_1-e_1^*;\cr
[e_2,e_2^*]=r_{22}e_2-e_2^*. \cr}\right.\\
&&\mbox{}\hspace{1.3cm}\bigcup \{\left(\matrix{r_{11}&r_{11}\cr
r_{11}& r_{11}\cr}\right)|r_{11}\ne 0\} \Longrightarrow
\left\{\matrix{[e_1, e_1^*]=r_{11}e_1+r_{11}e_2-e_1^*;\cr
[e_2,e_2^*]=
r_{11}e_1+r_{11}e_2-e_2^*;\cr [e_1^*, e_2^*]=r_{11}e_1^*-r_{11}e_2^*.\cr}\right.\\
&& {\rm SE(AII)}=\{\left(\matrix{0&r_{12}\cr
r_{12}&r_{22}\cr}\right)\} \Longrightarrow
\left\{\matrix{[e_1,e_1^*]=r_{12}e_2-e_1^*;\cr
[e_2,e_2^*]=r_{12}e_2-e_2^*;\cr
[e_1,e_2^*]=r_{12}e_1+r_{22}e_2-e_2^*.\cr}\right.\\
&&\mbox{}\hspace{1.3cm}\bigcup \{\left(\matrix{r_{11}&0\cr
0&0\cr}\right)|r_{11}\ne 0\} \Longrightarrow
\left\{\matrix{[e_1,e_1^*]=r_{11}e_1-e_1^*; &[e_1, e_2^*]=-e_2^*;\cr
[e_2,e_2^*]=r_{11}e_1-e_1^*;&[e_1^*e_2^*]=-r_{11}e_2^*.\cr}\right.\\
&&{\rm SE(AIII)}=\{\left(\matrix{r_{11}&0\cr 0&r_{22}\cr}\right)\}
\Longrightarrow
\left\{\matrix{[e_1,e_1^*]=r_{11}e_1-e_1^*\cr}.\right.\\
&&{\rm SE(AIV)}=\{\left(\matrix{r_{11}&r_{12}\cr
r_{12}&r_{22}\cr}\right)\} \Longrightarrow
\left\{\matrix{[e_1,e_2]=[e_1,e_1^*]=[e_2,e_2^*]=0;\cr [e_1,e_2^*]=[e_2,e_1^*]=[e_1^*,e_2^*]=0. \cr}\right.\\
&&{\rm SE(AV)}=\{\left(\matrix{0&r_{12}\cr
r_{12}&r_{22}\cr}\right)\} \Longrightarrow
\left\{\matrix{[e_1,e_1^*]=r_{12}e_2-e_1^*.\cr} \right.\\
&&{\rm SE(NI)}=\{\left(\matrix{r_{11}&\pm\sqrt{r_{11}r_{22}}\cr
\pm\sqrt{r_{11}r_{22}}&r_{22}\cr}\right)\} \Longrightarrow
\left\{\matrix{[e_1,e_2]=e_2;\cr [e_1,e_2^*]=r_{22}e_1;\cr
[e_1,e_1^*]=\pm\sqrt{r_{11}r_{22}}e_1;\cr
[e_2,e_1^*]=-2r_{11}e_1\mp\sqrt{r_{11}r_{22}}e_2+e_1^*;\cr
[e_2,e_2^*]=\mp\sqrt{r_{11}r_{22}}e_1-r_{22}e_2+e_2^*;\cr [e_1^*,e_2^*]=\mp\sqrt{r_{11}r_{22}}e_2^*-r_{22}e_1^*.\cr}\right.\\
&&{\rm SE(NII}_k,k\ne {\pm 1})=\{\left(\matrix{0&0\cr
0&r_{22}\cr}\right)\} \Longrightarrow \left\{\matrix{[e_1,e_2]=e_1;
\quad [e_1,e_2^*]=r_{22}e_1\cr [e_2,e_2^*]=e_1^*;\quad [e_1^*,
e_2^*]=-r_{22}e_1^*;\cr
[e_2, e_2^*]=kr_{22}e_2-ke_2^*.\cr}\right.\\
&&{\rm SE(NII}_1)=\{\left(\matrix{r_{11}&r_{12}\cr
r_{12}&0\cr}\right)\} \Longrightarrow
\left\{\matrix{[e_1,e_2]=e_1;\quad [e_1,e_1^*]=r_{12}e_1;\cr
[e_2,e_1^*]=-2r_{11}e_1-r_{12}e_2+e_1^*;\cr [e_2,e_2^*]=-e_2^*;\quad
[e_1^*,e_2^*]=-r{12}e_2^*.\cr}\right.\\
&&\mbox{}\hspace{1.5cm}\bigcup \{\left(\matrix{0&0\cr
0&r_{22}\cr}\right)|r_{22}\ne 0\} \Longrightarrow
\left\{\matrix{[e_1,e_2]=e_1;\quad [e_1,e_2^*]=r_{22}e_1;\cr
[e_2,e_1^*]=e_1^*;\quad [e_1^*,e_2^*]=-r_{22}e_1^*;\cr
[e_2,e_2^*]=r_{22}e_2-e_2^*.\cr}\right.\\
&&{\rm SE(NII}_{-1})=\{\left(\matrix{r_{11}&0\cr 0&0\cr}\right)\}
\Longrightarrow
\left\{\matrix{[e_1,e_2]=e_1;\quad[e_2,e_2^*]=e_2^*;\cr
[e_2,e_1^*]=-2r_{11}e_1+e_1^*-e_2^*.\cr}\right.\\
&&{\rm SE(NIII)}=\{\left(\matrix{r_{11}&\pm\sqrt{r_{11}r_{22}}\cr
\pm\sqrt{r_{11}r_{22}} &r_{22}\cr}\right)\} \Longrightarrow
\left\{\matrix{[e_1,e_2]=e_1;\quad [e_1,e_2^*]=r_{22}e_1;\cr
 [e_1,e_1^*]=\pm2\sqrt{r_{11}r_{22}}e_1+r_{22}e_2-e_2^*;\cr
 [e_2,e_1^*]=-r_{11}e_1;\quad
 [e_2,e_2^*]=r_{22}e_2-e_2^*.\cr}\right.\\
 &&{\rm SE(NIV}_k,k\ne
 {0,2})=\{\left(\matrix{r_{11}&(1-k)r_{11}\cr
 (1-k)r_{11}&(1-k)^2r_{11}\cr}\right)\}\\
&&\mbox{}\hspace{3cm} \Longrightarrow
 \left\{\matrix{[e_1, e_2]=e_1;\quad [e_1,e_2^*]=(1-k)^2r_{11}e_1;\cr
 [e_1,e_1^*]=(1-k)^2r_{11}e_1+k(1-k)^2r_{11}e_2-ke_2^*;\cr
 [e_2,e_1^*]=(1-k)e_1^*-r_{11}e_1-e_2^*;\cr
 [e_2,e_2^*]=-(1-k)r_{11}e_1+k(1-k)r_{11}e_2-ke_2^*;\cr[e_1^*,e_2^*]=-(1-k)^3r_{11}e_1^*+k(1-k)^2e_2^*.\cr}\right.\\
&&{\rm SE(NIV)}_2=\{\left(\matrix{r_{11}&-r_{12}\cr
-r_{12}&r_{22}\cr}\right)\} \Longrightarrow \left\{\matrix{[e_1,
e_2]=e_1;\quad [e_1,e_2^*]=r_{22}e_1;\cr
[e_1,e_1^*]=-3r_{22}e_1+2r_{22}e_2-2e_2^*;\cr
[e_2,e_1^*]=-r_{22}e_1-e_1^*-e_2^*;\cr
[e_1^*,e_2^*]=r_{22}(e_1^*+e_2^*);\cr
[e_2,e_2^*]=-r_{22}e_1+2r_{22}e_2-2e_2^*.\cr}\right.\\
&&{\rm SE(NV)}=\{\left(\matrix{r_{11}&0\cr0&0\cr}\right)\}
\Longrightarrow \left\{\matrix{[e_1,e_2]=e_2;\quad [e_1,
e_2^*]=-e_2^*;\quad [e_1^*,e_2^*]=-r_{11}e_2^*;\cr
[e_1,e_1^*]=2r_{11}e_1-2e_1^*;\quad
[e_2,e_1^*]=-r_{11}e_2-e_2^*.\cr}\right.\\
 &&\mbox{}\hspace{1.5cm}\bigcup \{\left(\matrix{r_{11}&0\cr
0&2r_{11}\cr}\right)|r_{11}\ne 0\} \Longrightarrow
\left\{\matrix{[e_1,e_2]=e_2;\quad [e_2,e_1^*]=r_{11}e_2-e_2^*;\cr
[e_1^*,e_2^*]=-r_{11}e_2^*;\quad [e_1,e_1^*]=2r_{11}-2e_1^*;\cr
[e_1, e_2^*]=4r_{11}e_2-e_2^*.\cr}\right.\\
&&\mbox{}\hspace{1.5cm}\bigcup \{\left(\matrix{r_{11}&-ir_{11}\cr
-ir_{11}&-r_{11}\cr}\right)|r_{11}\ne 0,i^2=-1\} \Longrightarrow
\left\{\matrix{[e_1,e_2]=e_2;\quad[e_2,e_2^*]=ir_{11}e_2;\cr
[e_1,e_1^*]=2r_{11}e_1-3ir_{11}e_2-2e_1^*;\cr
[e_1,e_2^*]=-ir_{11}e_1-2r_{11}e_2-e_2^*;\cr
[e_2,e_1^*]=-ir_{11}e_1-2r_{11}e_2-e_2^*;\cr
[e_1^*,e_2^*]=-2ir_{11}e_1^*-2r_{11}e_2^*.\cr}\right.
\end{eqnarray*}}

{\bf Corollary 4.1}\quad There are the invertible solutions of
$S$-equation in the following 2-dimensional left-symmetric algebras:
(AI)-(AV), (NII)$_1$, (NIV)$_2$, (NV).

{\bf Remark 4.2}\quad In fact, the 4-dimensional symplectic Lie
algebras were described in [MR1].

\section{Some 6-dimensional phase spaces}

The complete classification of 3-dimensional complex left-symmetric
algebras is very complicated ([Ba4]). On the other hand, it is
interesting to consider the solutions of $S$-equation in simple
left-symmetric algebras (without any ideal besides zero and itself),
like considering the solution of classical Yang-Baxter equation in
semisimple Lie algebras ([BD]). Although there is not a complete
classification of simple left-symmetric algebras, either, we have
known that (NV) is the only 2-dimensional simplex left-symmetric
algebra and the classification of 3-dimensional complex simplex
left-symmetric algebras is given as follows ([Bu],[Ba1], [Ba4]).

${\rm T}_1^{\lambda}=\langle e_1,e_2,e_3|e_1\cdot
e_1=(\lambda+1)e_1,e_1\cdot e_2=e_2,e_1\cdot e_3=\lambda
e_3,e_2\cdot e_3=e_3\cdot e_2=e_1\rangle $;

\mbox{}\hspace{1cm} $0\langle |\lambda|\langle 1,\;{\rm or}\;
\lambda=e^{i\theta},0\leq\theta\leq \pi$;

${\rm T}_2=\langle e_1,e_2,e_3|e_1\cdot e_1=\frac{3}{2}e_1,e_1\cdot
e_2=e_2,e_1\cdot e_3=\frac{1}{2} e_3,e_2\cdot e_3=e_3\cdot
e_2=e_1,e_3\cdot e_3=-e_2\rangle $.

Similarly as the discussion in the above section, by a direct
computation, we give the solutions of $S$-equation in the above
(3-dimensional) simple left-symmetric algebras and their
corresponding (6-dimensional) phase spaces (only the non-zero
products are given) as follows.

{\small \begin {eqnarray*} &&{\rm SE(T}_1^\lambda,\lambda\ne
1)=\{\left(\matrix{r_{11}&0&0\cr 0&0&0\cr 0&0&0\cr}\right)\}
\Longrightarrow \left \{\matrix{[e_1,e_2]=e_2;\;\; [e_1,e_3]=\lambda
e_3;\cr [e_1,e_1^*]=(\lambda+1)r_{11}e_1-(\lambda+1)e_1^*;\cr
[e_1,e_2^*]=-e_2^*;\;\; [e_1,e_3^*]=-\lambda e_3^*;\cr
[e_2,e_1^*]=-r_{11}e_2-e_3^*;\;\; [e_1^*,e_2^*]=-r_{11}e_2^*;\cr
[e_1^*,e_3^*]=-\lambda r_{11}e_3^*;\;\; [e_3,e_1^*]=-\lambda
r_{11}e_3-e_2^*.\cr}\right.\\
&&\mbox{}\hspace{1.5cm}\bigcup \{\left(\matrix{0&0&0\cr
0&r_{22}&0\cr 0&0&0\cr}\right)|r_{22}\ne 0 \} \Longrightarrow
\left\{\matrix{[e_1,e_2]=e_2;\quad [e_1,e_3]=\lambda e_3;\cr
[e_1,e_1^*]=-(\lambda+1)e_1^*;\quad [e_1,e_2^*]=2r_{22}e_2-e_2^*;\cr
[e_1,e_3^*]=-\lambda e_3^*;\quad [e_2,e_1^*]=-e_3^*;\cr
[e_1^*,e_2^*]=r_{22}e_3^*;\quad [e_3,e_1^*]=r_{22}e_2-e_2^*.\cr}\right.\\
&&\mbox{}\hspace{1.5cm}\bigcup \{\left(\matrix{0&0&0\cr 0&0&0\cr
0&0&r_{33}\cr}\right)|r_{33}\ne 0\} \Longrightarrow
\left\{\matrix{[e_1,e_2]=e_2;\quad [e_1,e_3]=\lambda e_3;\cr
[e_1,e_1^*]=-(\lambda+1)e_1^*;\quad [e_1,e_2^*]=-e_2^*;\cr
[e_1,e_3^*]=2\lambda r_{33}e_3-\lambda e_3^*;[e_3,e_1^*]=-e_2^*;\cr
[e_2,e_1^*]=r_{33}e_3-e_3^*;\quad [e_1^*,e_3^*]=r_{33}e_3^*.\cr}\right.\\
&&\mbox{}\hspace{1.5cm}\bigcup \{\left(\matrix{r_{11}&0&0\cr
0&0&(\lambda+1)r_{11}\cr
0&(\lambda+1)r_{11}&0\cr}\right)|(\lambda+1)r_{11}\ne 0\}\\
&&\mbox{}\hspace{3cm} \Longrightarrow \left
\{\matrix{[e_1,e_2]=e_2;\quad[e_1,e_3]=\lambda e_3;\cr
[e_1,e_1^*]=(\lambda+1)r_{11}e_1-(\lambda+1)e_1^*;\cr
[e_1,e_2^*]=(\lambda+1)^2r_{11}e_3-e_2^*;\cr
[e_1,e_3^*]=(\lambda+1)^2r_{11}e_2-\lambda e_3^*;\cr
[e_1^*,e_2^*]=\lambda r_{11}e_2^*;\quad
[e_3,e_1^*]=r_{11}e_3-e_2^*;\cr [e_2,e_1^*]=\lambda
r_{11}e_2-e_3^*;\quad [e_1^*,e_3^*]=r_{11}e_3^*.\cr}\right.\\
&&{\rm
SE( T}_1^1)={\rm SE(T}_1^\lambda,\;\;{\rm set}\;\;\lambda=1)\\
&&\mbox{}\hspace{1.5cm}\bigcup
\{\left(\matrix{\sqrt{r_{22}r_{33}}&0&0\cr
0&r_{22}&\sqrt{r_{22}r_{33}}\cr
0&\sqrt{r_{22}r_{33}}&r_{33}\cr}\right)| r_{22}r_{33}\ne0\}\\
&&\mbox{}\hspace{3cm}\Longrightarrow
\left\{\matrix{[e_1,e_2]=e_2;\quad[e_1,e_3]=e_3;\quad
[e_2,e_1^*]=r_{33}e_3-e_3^*;\cr[e_1,e_1^*]=2\sqrt{r_{22}r_{33}}e_1-2e_1^*;\cr
[e_1,e_2^*]=2r_{22}e_2+2\sqrt{r_{22}r_{33}}e_3-e_2^*;\cr
[e_1^*,e_2^*]=r_{22}e_3^*;\quad[e_1^*,e_3^*]=r_{33}e_2^*;\quad
[e_3,e_1^*]=r_{22}e_2-e_2^*;\cr[e_1e_3^*]=2\sqrt{r_{22}r_{33}}e_2+2r_{33}e_3-e_3^*.\cr
}\right.\\
&&\mbox{}\hspace{1.5cm}\bigcup\{\left(\matrix{-i\sqrt{2r_{12}r_{13}}&r_{12}&r_{13}\cr
r_{12}&ir_{12}\sqrt{\frac{r_{12}}{2r_{13}}}&i\sqrt{\frac{r_{12}r_{13}}{2}}\cr
r_{13}&i\sqrt{\frac{r_{12}r_{13}}{2}}&ir_{13}\sqrt{\frac{r_{13}}{2r_{12}}}\cr}\right)|r_{12}r_{13}\ne0
\}\\
&&\mbox{}\hspace{3cm} \Longrightarrow \left
\{\matrix{[e_1,e_2]=e_3;\quad [e_1,e_3]=e_3;\cr
{\small[e_1,e_1^*]=-2i\sqrt{2r_{12}r_{13}}e_1+3r_{12}e_2+3r_{12}e_2+3r_{13}e_3-2e_1^*};\cr
[e_1,e_2^*]=r_{12}e_1+2ir_{12}\sqrt{\frac{r_{12}}{2r_{13}}}e_2+i\sqrt{2r_{12}r_{13}}e_3-e_2^*;\cr
[e_1,e_3^*]=r_{13}e_1+i\sqrt{2r_{12}r_{13}}e_2+2ir_{13}\sqrt{\frac{r_{13}}{2r_{12}}}e_3-e_3^*;\cr
[e_2,e_1^*]=r_{13}e_1+\frac{3i}{2}\sqrt{r_{12}r_{13}}+ir_{13}\sqrt{\frac{r_{13}}{2r_{12}}}e_3-e_3^*;\cr
[e_2,e_2^*]=-r_{12}e_2;\quad [e_2,e_3^*]=-r_{13}e_2;\quad
[e_3,e_2^*]=-r_{12}e_3;\cr
[e_3,e_1^*]=r_{12}e_1+ir_{12}\sqrt{\frac{r_{12}}{2r_{13}}}e_2+\frac{3i}{2}\sqrt{2r_{12}r_{13}}e_3-e_2^*;\cr
[e_3,e_3^*]=-r_{13}e_3;\quad
[e_2^*,e_3^*]=r_{13}e_2^*-r_{12}e_3^*;\cr
[e_1^*,e_2^*]=2r_{12}e_1^*+\frac{3i}{2}\sqrt{2r_{12}r_{13}}e_2^*+ir_{12}\sqrt{\frac{r_{12}}{2r_{13}}}e_3^*;\cr
[e_1^*,e_3^*]=2r_{13}e_1^*+ir_{13}\sqrt{\frac{r_{13}}{2r_{12}}}e_2^*+\frac{3i}{2}\sqrt{2r_{12}r_{13}}e_3^*.\cr}\right.\\
&&\mbox{}\hspace{1.5cm}\bigcup
\{\left(\matrix{-\sqrt{r_{22}r_{33}}&0&0\cr
0&r_{22}&-\sqrt{r_{22}r_{33}}\cr
0&-\sqrt{r_{22}r_{33}}&r_{33}\cr}\right)| r_{22}r_{33}\ne0\}\\
&&\mbox{}\hspace{3cm}\Longrightarrow
\left\{\matrix{[e_1,e_2]=e_2;\quad[e_1,e_3]=e_3;\quad
[e_1,e_1^*]=-2\sqrt{r_{22}r_{33}}e_1-2e_1^*;\cr
[e_2,e_1^*]=r_{33}e_3-e_3^*;\quad
[e_1,e_2^*]=2r_{22}e_2-2\sqrt{r_{22}r_{33}}e_3-e_2^*;\cr
[e_1^*,e_2^*]=r_{22}e_3^*;\quad [e_1^*,e_3^*]=r_{33}e_2^*;\quad
[e_3,e_1^*]=r_{22}e_2-e_2^*;\cr
[e_1,e_3^*]=-2\sqrt{r_{22}r_{33}}e_2+2r_{33}e_3-e_3^*.\cr}\right.\\
&&\mbox{}\hspace{1.5cm}\bigcup
\{\left(\matrix{i\sqrt{2r_{12}r_{13}}&r_{12}&r_{13}\cr
r_{12}&-ir_{12}\sqrt{\frac{r_{12}}{2r_{13}}}&-i\sqrt{\frac{r_{12}r_{13}}{2}}\cr
r_{13}&-i\sqrt{\frac{r_{12}r_{13}}{2}}&-ir_{13}\sqrt{\frac{r_{13}}{2r_{12}}}\cr}\right)
\}\\
&&\mbox{}\hspace{3cm} \Longrightarrow \left
\{\matrix{[e_1,e_2]=e_3;\quad [e_1,e_3]=e_3;\cr
[e_1,e_1^*]=2i\sqrt{2r_{12}r_{13}}e_1
+3r_{12}e_2+3r_{13}e_3-2e_1^*;\cr
[e_1,e_2^*]=r_{12}e_1-2ir_{12}\sqrt{\frac{r_{12}}{2r_{13}}}e_2
-i\sqrt{2r_{12}r_{13}}e_3-e_2^*;\cr
[e_1,e_3^*]=r_{13}e_1-i\sqrt{2r_{12}r_{13}}e_2-2ir_{13}\sqrt{\frac{r_{13}}{2r_{12}}}e_3-e_3^*;\cr
[e_2,e_1^*]=r_{13}e_1-\frac{3i}{2}\sqrt{r_{12}r_{13}}-ir_{13}\sqrt{\frac{r_{13}}{2r_{12}}}e_3-e_3^*;\cr
[e_2,e_2^*]=-r_{12}e_2;\quad [e_2,e_3^*]=-r_{13}e_2;\quad
[e_3,e_2^*]=-r_{12}e_3;\cr
[e_3,e_1^*]=r_{12}e_1-ir_{12}\sqrt{\frac{r_{12}}{2r_{13}}}e_2-\frac{3i}{2}\sqrt{2r_{12}r_{13}}e_3-e_2^*;\cr
[e_3,e_3^*]=-r_{13}e_3;\quad
[e_2^*,e_3^*]=r_{13}e_2^*-r_{12}e_3^*;\cr
[e_1^*,e_2^*]=2r_{12}e_1^*-\frac{3i}{2}\sqrt{2r_{12}r_{13}}e_2^*-ir_{12}\sqrt{\frac{r_{12}}{2r_{13}}}e_3^*;\cr
[e_1^*,e_3^*]=2r_{13}e_1^*-ir_{13}\sqrt{\frac{r_{13}}{2r_{12}}}e_2^*-\frac{3i}{2}\sqrt{2r_{12}r_{13}}e_3^*.\cr}\right.\\
&&{\rm SE(T}_2)=\{\left(\matrix{r_{11}&0&0\cr 0&0&0\cr
0&0&0\cr}\right)\} \Longrightarrow \left
\{\matrix{[e_1,e_2]=e_2;\quad [e_1,e_3]=\frac{1}{2}e_3;\cr
[e_1,e_1^*]=\frac{3}{2}r_{11}e_1-\frac{3}{2}e_1^*;\quad
[e_1,e_2^*]=-e_2^*;\cr
[e_1,e_3^*]=-\frac{1}{2}e_3^*;\quad[e_2,e_1^*]=-r_{11}e_2-e_3^*;\cr
[e_3,e_2^*]=e_3^*;\quad[e_1^*,e_2^*]=-r_{11}e_2^*;\cr
[e_3^*,e_1^*]=\frac{r_{11}}{2}e_3^*;\quad
[e_1^*,e_3]=\frac{r_{11}}{2}e_3-e_2^*.\cr}\right.\\
&&\mbox{}\hspace{1.5cm}\bigcup \{\left(\matrix{0&0&0\cr
0&r_{22}&0\cr 0&0&0\cr}\right)|r_{22}\ne0\} \Longrightarrow
\left\{\matrix{[e_1,e_2]=e_2;\quad [e_1,e_3]=\frac{1}{2}e_3;\cr
[e_1,e_1^*]=-\frac{3}{2}e_1^*;\quad [e_1^*,e_2^*]=-r_{22}e_3^*\cr
[e_1,e_3^*]=-\frac{1}{2}e_3^*;\quad[e_2,e_1^*]=-e_3^*;\cr
[e_3,e_1^*]=r_{22}e_2-e_2^*;\quad[e_3,e_2^*]=e_3^*;\cr
[e_1,e_2^*]=2r_{22}e_2-e_2^*.\cr}\right.\\
&&\mbox{}\hspace{1.5cm}\bigcup \{\left(\matrix{r_{11}&0&0\cr
0&0&\frac{3}{2}r_{11}\cr
0&\frac{3}{2}r_{11}&0\cr}\right)|r_{11}\ne0\} \\
&&\mbox{}\hspace{3cm}\Longrightarrow
\left\{\matrix{[e_1,e_2]=e_2;\quad [e_1,e_3]=\frac{1}{2}e_3;\quad
[e_1,e_1^*]=\frac{3}{2}r_{11}e_1-\frac{3}{2}e_1^*;\cr
[e_1,e_2^*]=\frac{9}{4}r_{11}e_3-e_2^*;\quad
[e_1,e_3^*]=\frac{9}{4}r_{11}e_1-\frac{1}{2}e_3^*;\cr
[e_2,e_1^*]=\frac{1}{2}r_{11}e_2-e_3^*;\quad
[e_3,e_1^*]=r_{11}e_3-e_2^*;\cr [e_1^*,e_2^*]=\frac{1}{2}e_2^*;\quad
[e_1^*,e_3^*]=r_{11}e_3^*;\quad
[e_3,e_2^*]=-\frac{3}{2}r_{11}e_2+e_3^*.\cr}\right.
\end {eqnarray*}}

{\bf Corollary 5.1}\quad There are invertible solutions of
$S$-equation in 3-dimensional simple left-symmetric algebras besides
(T$_1^{-1}$).

\section{Conclusion and discussion}

From the study in the previous sections, we would like to give the
following conclusion and discussion.

(1) The phase space of a Lie algebra obtained from a non-zero
symmetric solution of $S$-equation in a left-symmetric algebra is
symplectically isomorphic to the one obtained from the zero
solution. Therefore, it is invalid to obtain the new symplectic Lie
algebras from this way. Moreover, there is a quite similar property
of classical Yang-Baxter equation which was given in [DiM]. It is
interesting to know that both of them are interpreted in terms of
left-symmetric algebras.

(2) We have obtained the phase spaces from all solutions of
$S$-equation in 2-dimensional complex left-symmetric algebras and
3-dimensional complex simple left-symmetric algebras. Comparing with
the construction in [Ku1], these phase spaces (parak\"ahler Lie
algebras) are new and different. Furthermore, we would like to point
out that it is hard and less practicable to extend what we have done
in section 4 and 5 to other cases in higher dimensions since the
$S$-equation in left-symmetric algebras involves the (nonlinear)
quadratic equations (4.1).

(3) We have not proved yet whether the phase spaces obtained from
$S$-equation in left-symmetric algebras are not isomorphic for any
two different parameters as parak\"ahler Lie algebras. In fact, they
are closely related to ${\rm SE}(A)$ of a left-symmetric algebra $A$
which relays on the choice of a basis of $A$ and its corresponding
structural constants. It is natural to consider whether there is  a
meaningful ``classification rules" so that the classification of the
solutions of $S$-equation in left-symmetric algebras can be more
``interesting"?

(4) Our study on phase spaces in low dimensions can provide some
good examples to study certain related geometric structures, like
complex and K\"ahler structures ([Ba2]). It is natural to consider
the possible application in physics. Furthermore, we hope that they
can be a guide for the cases in higher dimensions, even in infinite
dimension. For example, since left-symmetric algebras are the
underlying spaces of vertex algebras which is the algebraic
structure of conformal field theory ([BK]) and play a crucial role
in the Hopf algebraic approach of Connes and Kreimer to
renormalization theory of perturbative quantum field theory ([CK]),
it would be interesting to consider the roles of phase spaces and
$S$-equation related to the left-symmetric algebras there.

\section*{Acknowledgements}

The authors thank Xiaoli Kong for the valuable discussion and the
referees' important suggestion. This work was supported in part by
the National Natural Science Foundation of China (10571091,
10621101), NKBRPC (2006CB805905)£¬ and Program for New Century
Excellent Talents in University.

\end{document}